\documentclass[11pt]{article}

\usepackage[T1]{fontenc}              % T1 font encoding as default
\usepackage[latin9]{inputenc}         % ISO-8859-15 (Latin 9) text

\usepackage[a4paper]{geometry}        % DIN A4 paper

\usepackage{url}

\usepackage{booktabs}

\usepackage{graphicx} % WE USE PDFLATEX
\usepackage{amssymb,amsfonts,amsmath}

\usepackage{natbib}
\usepackage{mathpazo}
\usepackage{microtype}

\usepackage{color}

\usepackage{todonotes}
 
\usepackage{multirow}

\newcommand{\segemehl}{{\tt segemehl} }

\newcommand{\figcite}[1]{Fig.~\textbf{\ref{#1}}}
\newcommand{\eqcite}[1]{Eq.~\textbf{\ref{#1}}}

\begin{document}

\date{19 May 2014; revised 10 January 2015}

\title{
A Simple Data-Adaptive \\Probabilistic Variant Calling Model
}

\newcommand\myaddress[2][]{\relax
    {\noindent{\small$^{#1}$#2}}}

\author{Steve Hoffmann$^{1,2,3}$ 
\thanks{To whom correspondence should be addressed. Email: {\tt steve@bioinf.uni-leipzig.de} } 
{}, Peter F. Stadler$^{2-8}$, and Korbinian Strimmer$^{9,10}$ 
}

\maketitle

\myaddress[1]{Junior Research Group Transcriptome Bioinformatics, 
  University Leipzig, H{\"a}rtelstra{\ss}e
  16-18, Leipzig, Germany\\}
\myaddress[2]{Interdisciplinary Center for Bioinformatics and Bioinformatics 
  Group, University Leipzig, H{\"a}rtelstra{\ss}e 16-18, Leipzig, Germany\\}
\myaddress[3]{LIFE Research Center for Civilization Diseases, University 
  Leipzig, H{\"a}rtelstra{\ss}e 16-18, Leipzig, Germany\\}
\myaddress[4]{Department of Theoretical Chemistry, University Vienna, 
  W{\"a}hringer Stra{\ss}e 17, Vienna, Austria\\}
\myaddress[5]{RNomics Group, Fraunhofer Institute for Cell Therapy and 
  Immunology -- IZI, Perlickstra{\ss}e 1, Leipzig, Germany\\}
\myaddress[6]{Max-Planck-Institute for Mathematics in the Sciences, 
  Inselstra{\ss}e 22, Leipzig, Germany\\}
\myaddress[7]{Center for non-coding RNA in Technology and 
  Health, University of Copenhagen, Gr{\o}nneg\r{a}rdsvej 3, 
  Frederiksberg, Denmark\\}
\myaddress[8]{Santa Fe Institute, 1399 Hyde Park Road, 
  Santa Fe, USA\\}
\myaddress[9]{Institute for Medical Informatics, Statistics and Epidemiology,
  University of Leipzig, H{\"a}rtelstra{\ss}e 16--18, D-04107 Leipzig, 
  Germany\\}
\myaddress[\textsuperscript{10}]{Dept. of Epidemiology and Biostatistics, 
  Imperial College London, Norfolk Place, London W2 1PG, UK.}

\newpage

\definecolor{green}{gray}{0} % redefine green as black (don't show changes in ms)

\begin{abstract}

\noindent\textbf{Background:} Several sources of noise obfuscate the
identification of single nucleotide variation (SNV) in next generation
sequencing data. { For instance, errors may be introduced
during library construction and sequencing steps. In addition, the
reference genome and the algorithms used for the alignment of the
reads are further critical factors determining the efficacy of variant
calling methods.  It is crucial to account for these factors
in individual sequencing experiments.}

\noindent\textbf{Results:} We introduce a simple data-adaptive model
for variant calling. { This model automatically adjusts
to specific factors such as alignment errors}. To achieve this,
several characteristics are sampled from sites with low mismatch
rates, and { these are used} to estimate empirical
log-likelihoods. These likelihoods are then combined to a score that
typically gives rise to a mixture distribution. From these we
determine a decision threshold to separate potentially variant sites
from the noisy background.

\noindent\textbf{Conclusions:} In simulations we show that our simple
proposed model is competitive with frequently used much more complex
SNV calling algorithms in terms of sensitivity and
specificity. { It performs specifically well in cases
with low allele frequencies.} The application to next-generation
sequencing data reveals stark differences of the score distributions
indicating a strong influence of data specific sources of noise. The
proposed model is specifically designed to adjust to these
differences.

\end{abstract}

\newpage

\section*{Background}

Recent studies report a strikingly low concordance of currently available
methods and pipelines for identification of single nucleotide variation (SNV), 
both somatic and germline, indicating that both computational methods as well as 
sequencing protocols have a major impact on the sensitivity and specificity
of the variation calling tool~\citep{Rawe2013}. Specifically, the allelic
fraction as well as the coverage of the variant allele are crucial
determinants for the statistical benchmarks~\citep{Xu2014, Yu2013}.
Practical guidelines of SNV callers such as GATK~\citep{MH+2010} or
SAMtools~\citep{LH+2009} suggest to apply rigorous postprocessing filters
to reduce the number of false positive calls. Other studies indicate that
the application of these filters {leads} to a substantial improvement of the
concordance of the callers~\citep{Liu2013}.  Nevertheless, applying
stringent thresholds for variables such as the strand bias, the coverage or
read start variation bears the risk of losing important
information~\citep{Pabinger2013}. These authors emphasize that the
different { algorithmic} and statistical components of a variant caller have
to be evaluated as a whole and cannot not be meaningfully judged as single
components.

If DNA library preparation protocols and sequencing machines
were able to produce error-free and unbiased sequences of sufficient length
the task of variant calling would be easy.  Due to various error sources
and technical limitations of library preparation, sequencing, and
alignment, however, a substantial level of noise complicates the analysis.
Since these factors can not be totally controlled during the experiment it
seems reasonable to adjust the thresholds for calling a variant depending
on the separability of noise and signal, i.e. the true variants. During
amplification incorrect nucleotides are incorporated with some error rate
$\epsilon_f$ and during the sequencing step incorrect nucleotides are
called with the rate $\epsilon_g$. After the alignment of the reads to a
reference sequence we may observe these errors as mismatches or
indels. Additional mismatches and indels are caused by these reference and
alignment errors ($\epsilon_a$).  The mismatch rate of a genomic site can
be assumed to be the sum $\delta = \epsilon + \beta$, where $\beta$
represents the biological variation and $\epsilon$ is the compound effect
of the technical errors $\epsilon_f$, $\epsilon_g$, and $\epsilon_a$.
\figcite{fig:pcrclones} summarizes this situation.

The two most commonly used tools for SNV calling methods, SAMtools
and GATK, employ probabilistic models for variant calling. Specifically,
the algorithm used by SAMtools~\citep{Li2011a} is based on the
likelihood of a genotype which is computed as 
\begin{equation}
  L(g) = \frac{1}{m^k} 
     \prod^l_{j=1}   \lbrack(m-g)\zeta_j+g(1-\zeta_j)\rbrack 
     \prod^k_{j=l+1} \lbrack(m-g)(1-\zeta_j)+g\zeta_j\rbrack\,,
\label{eq:samtools}
\end{equation} 
where $g$ denotes the number of reference alleles, $m$ the ploidy, $k$ the
number of reads seen at a site, and $\zeta$ the error probability delivered
by the sequencer. \eqcite{eq:samtools} assumes that the first $l$ bases are
identical to the reference, the subsequent bases are not.  Subsequently,
from this a likelihood for the allele count $L(c)$ is obtained. Using the
observed allele frequency spectrum { $\phi_c$ as prior
  information a posterior probability
\begin{equation}
   \Pr\{ c \} = 
  \frac{\phi_c L(c)} {\sum_{c'} \phi_{c'} L({c'})}
\end{equation}
} is computed, and a variant is called if $\Pr\{ c>1 \}$ exceeds a
certain specified threshold.

The GATK pipeline uses a related probabilistic model for calling
variants~\citep{DePristo2011}.  Similar to SAMtools, the probability
$\Pr\{D_{j} | A\}$ of observing the base $D_{j}$ under the hypothesis that
$A$ is the true base is calculated by
\begin{equation}
  \Pr\{D_j | A\} = 
  \begin{cases}
    1 - \zeta_j & D_j = A\\
    \zeta_j \; \Pr\{\text{A is true} | \text{$D_j$ is miscalled}\} & 
    \text{otherwise,}
  \end{cases}
\end{equation}
where $\Pr\{\text{A is true} | \text{$D_j$ is miscalled}\}$ is a
precomputed, sequencer specific lookup table.  Using prior information
based on precomputed heterozygosity estimations GATK evaluates the
posterior probabilities of a site to be variant.  As with SAMtools calls
are determined using fixed preselected thresholds.

Here, we propose a simple probabilistic model for variant calling using a
data adaptive threshold on the scale of log-odd-ratios computed from
empirical distributions of certain site characteristics.  Our approach
allows to optimally separate simulated SNVs from the noisy background
without specification of a threshold for posterior probabilities.  In
brief, our model starts out by evaluating the mismatch frequencies $\delta$
in a data set. Subsequently, we sample several characteristics of the sites
with small $\delta$ to serve as empirical reference model.  The fundamental
idea used here is that the vast majority of sites is invariant and thus
allows to capture the features of the data specific error model.  These
characteristics are then used to form empirical log-likelihoods that are
combined to a log-odds type score. Typically, we observe a mixture
distribution of two score populations, which we may then separate by a decision
threshold.

Next, we discuss the details of our approach and the proposed data-adaptive
variant calling algorithm.  Subsequently, we apply our method to both
synthetic and next generation sequencing data from various species.
{ A} reference implementation in C99 of our method
{ called \texttt{haarz} is available at
  \url{http://www.bioinf.uni-leipzig.de/Software/}.}

\begin{figure}
\begin{center}
  \includegraphics{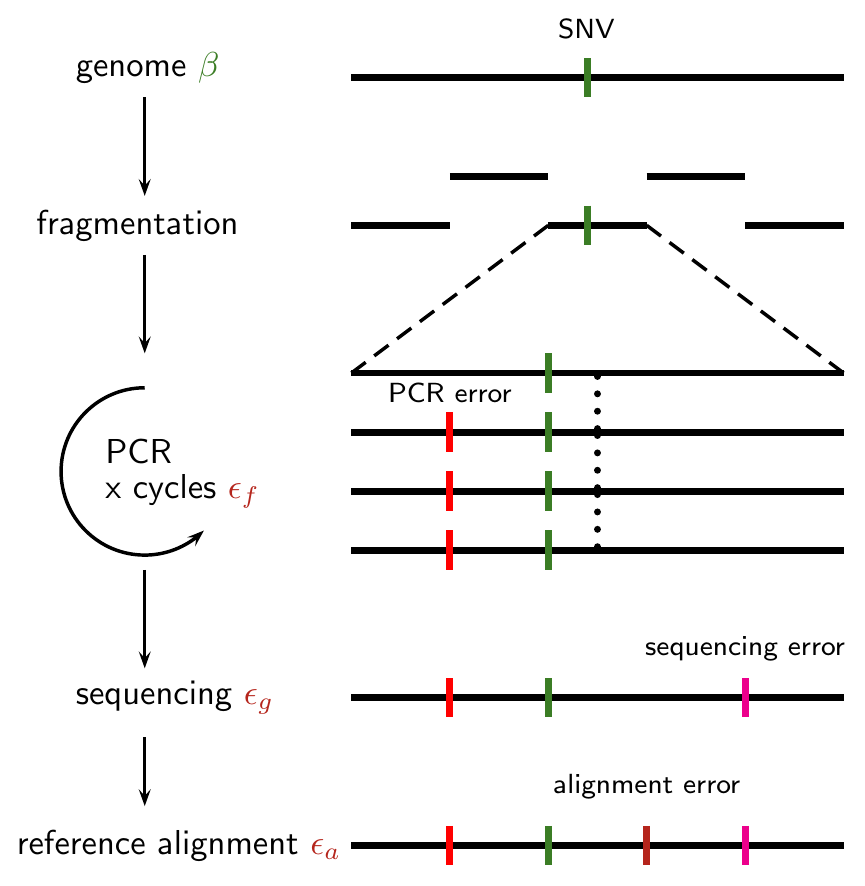}
\end{center}
\caption{\textbf{Accumulation and sources of errors in next generation
    sequencing reads.} The identification of single nucleotide variations
  (green dashes) is complicated by various sources of error. PCR errors
  accumulate during the amplification and sequencing step. After
  fragmentation single fragments undergo several amplification cycles.
  Errors are introduced with a rate $\delta_f$ (red dashes).  Further
  errors are accumulated during sequencing ($\delta_g$). During the
  alignment to the reference further mismatches and indels are introduced
  ($\delta_a$).\label{fig:pcrclones}}
\end{figure}

\section*{Methods} 

\subsection*{Notation}

We denote the position of a site in the reference genome by $i \in [1,
\ldots, n]$ where $n$ is the genome length.  After aligning the reads to a
genome each reference base typically has several nucleotides aligned to
it. We refer to the set of all aligned nucleotides as the \emph{cross
  section} $C^i$ at position $i$.  The coverage at position $i$ is the size
$|C^i|$ of this set.  We use the index $j \in [1, \ldots, |C^i|]$ to refer
to a specific read.  The length of read $j$ aligned to site $i$ is denoted
by $m^i_j$, and the position of a nucleotide in a read is denoted by $k^i_j
\in \{1, \ldots, m^i_j\}$.  For simplicity, we occasionally leave out the
index $i$ when there is no danger of ambiguity.

The nucleotide in a cross section can be partitioned into sets of match
($M$) and mismatch ($\overline{M}$) nucleotides so that $C^i = M^i \cup
\overline{M}^i$.  The variant calling algorithm described below uses the
partition $\{ M^i, \overline{M}^i \}$ at each position $i$ to compute an
overall score for this particular site.

\subsection*{Biological versus technical variation}

The \emph{mismatch rate} $\delta^i =|M^i| / |C^i|$ is the observed number
of mismatches divided by the coverage.  The mismatch rate $\delta^i =
\beta^i + \epsilon^i $ may be decomposed into biological and technical
variation, where $\epsilon^i$ denotes the technical error that accumulated
during the preparation, sequencing and alignment steps and $\beta^i$
denotes the biological nucleotide variation at site $i$.

We aim to distinguish biologically variant positions ($\beta^i > 0$) from
non-variant positions ($\beta^i = 0$), based on the observed mismatch rates
$\delta^i$ and site characteristic scores derived from sequence data or
produced during sequencing.

Cross sections with high mismatch rates are indicative of biological
variation in the sample, whereas in cross sections with small mismatch rate
the mismatches are more likely due to technical errors.  Conversely, in the
overwhelming majority of cross sections $C^i$ we may assume that there is
no biological variation present, i.e.\ $\beta^i=0$, and thus mismatches are
only caused by technical errors.

For use in the variant calling score we estimate for each  $\delta^i$ the
corresponding empirical quantile $q(\delta^i)$. The motivation for using the 
quantile rather than the actual value is that it implements a simple
normalization of the error.  The empirical quantile $q(\delta^i)$ is
estimated by tabulating the cumulative frequencies of $\delta^i$ across
the genome and then reading off the quantile from the resulting empirical
distribution function (ECDF).

To ascertain the probabilities of certain site characteristics,
discussed further below, we uniformly sample from sites with $0 < \delta^i <
0.5$. Informally, these characteristics then reflect ``background
distributions'' of non-variant sites and thus are estimated from 
sites with less than 50\% of mismatches. 

The degree of biological variation depends on the type of genome.  For
heterozygous genomes one expects to find predominantly SNP alleles with
$\beta^i=0.5$ or $\beta^i=1.0$, whereas cancer tissues may show mutations
with $0 < \beta^i \leq 1$ depending on the heterogeneity and cancer cell
content of the sample. Similarly, arbitrary values of $\beta_i$ will appear
in whole population sequencing data.  Accordingly, we expect different
values of $\beta^i$ for mixtures of sequencing data from different
individuals.  The variant calling algorithm introduced in the following
makes no assumptions concerning the presence of diploid genomes, knowledge
about the ploidy, homo- or heterozygosity.

\subsection*{Site characteristics}

In addition to the partitioning of nucleotides at a given site into match
and mismatch sets, our algorithm uses the following information, which { is}
typically reported by the sequencer or the read mapper for every site $i$
and read $j$:
\begin{itemize}
\item the nucleotide qualities ($Q$),
\item relative read position ($P$),
\item errors in the alignment ($R$), and
\item the number of multiple hits ($H$).
\end{itemize}
The nucleotide qualities take on values between $0$ and $1$ and are given
as $Q = 1 - \zeta$, i.e., as probability of a base being correct, with
values close to $1$ corresponding to optimally accurate sequencing.  We
directly use $Q$ in computing the variant calling score.

The relative read position is given by $P = k^i_j /m^i_j$.  For the
construction of our variant calling score we employ the probability
$\Pr(M|P)$ of a match at a given read position, along with the maximum
$P_{M} = \max_{P} \Pr(M|P)$.  The probability of a mismatch is then given
by $\Pr(\overline{M}|P) = 1 - \Pr(M|P)$, and its maximum $P_{\overline{M}}=
\max_{P} \Pr(\overline{M}|P)$.  We estimate the probability $\Pr(M|P)$
empirically, i.e., by appropriately counting matches and mismatches over
all sites and reads.

The { number of errors} in the alignment is an integer value greater or equal to zero,
and denoted here by $R$.  Finally, the number of multiple hits $H$
describes the number of alignments for each read. The multiplicity of an
alignment yields information on the repetitiveness of a genomic region.  As
above for the relative read position, we tabulate the occurrence of matches
for each value of $R$ and and $H$ and correspondingly obtain estimates of
the probabilities $\Pr(M|R)$ and $\Pr(M|H)$.

\subsection*{Scores for distinguishing variant and non-variant sites}

Informally, in a non-variant cross section ($\beta^i = 0$) we expect that
the probability of a match base increases with high nucleotide qualities
(good sequencing), low read error rates, few multiple hits and good read
positions.  Conversely, the probability of mismatching bases in non-variant
cross sections increases with low nucleotide qualities (poor sequencing),
high read error rates, multiple hits and error-prone read positions.  For
variant sites with $\beta > 0$ we expect to have high nucleotide qualities,
good read positions and few multiple hits also for the mismatch
bases. Consequently, for distinguishing variant from non-variant sites only
the mismatching bases are relevant.

We introduce four log-odds ratios to formalize and summarize the evidence 
for a variant over a non-variant based on the above four site
characteristics. 
$$
\Delta_{Q} = \frac{1}{|C|} \sum_{x\in\overline{M}} \log{Q_x\over 1-Q_x} 
$$
for base qualities,
$$
\Delta_{P} = \frac{1}{|C|} \sum_{x\in\overline{M}} 
   \left(\log{ \Pr(M|P_x) \over  1-\Pr(M|P_x) } + 
         \log{ P_{\overline{M}} \over P_{M} }    \right)
$$
for read positions, which are rescaled by their respective maximally
attained values $P_{M}$ and $P_{\overline{M}}$,
$$
\Delta_{R} = \frac{1}{|C|} \sum_{x\in\overline{M}}      
    \log{\Pr(M|R_x) \over 1-\Pr(M|R_x)} 
$$
for read errors, and
$$
\Delta_{H} = \frac{1}{|C|} \sum_{x\in\overline{M}}      
    \log{\Pr(M|H_x) \over 1-\Pr(M|H_x)} 
$$
for multiple matches. {Note that only reads with mismatching bases in a cross-section
are used for estimation, i.e. match bases are ignored.
If there are only match bases in a cross-section ,
i.e. if $| \overline{M} |=0$ then the cross-section {\it is not considered
in any component of our model}}.

\subsection*{Variant calling with adaptive threshold}

From these log-odds { ratios} we now construct a total score for variant
calling by computing, at any position $i$,
$$
S_i = \Delta_{P_i} + \Delta_{Q_i} + \Delta_{R_i} + \Delta_{H_i} + 
      \log q(\delta^i)\,. 
$$
This score comprises the four summaries of the site characteristics, as
well as the log-quantile of the observed mismatch rate $\delta^i$, i.e.
the observed number of changes at position normalized by coverage.  A low
quantile for $\delta^i$ thus strongly penalizes the overall score.

\begin{figure}[t]
\begin{center}
\includegraphics[scale=0.5]{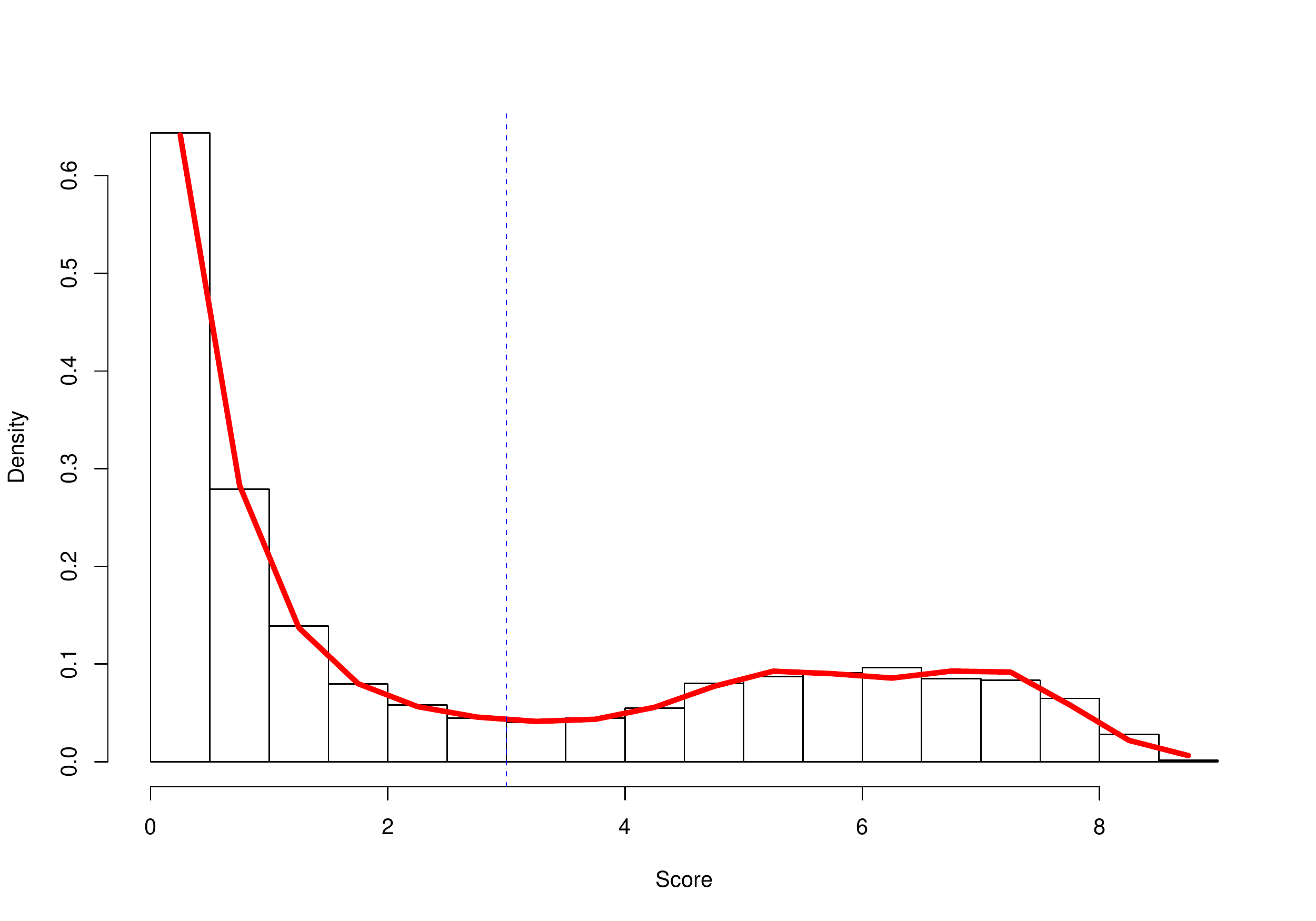}
\end{center}
\caption{\textbf{Adaptive cutoff determination.} The figure shows the
  distribution of the empirical scores $S_i>0$ (red continuous line) and
  the cutoff $S^*$ (dotted blue vertical line).  The threshold separates
  potential variants from background signal.\label{fig:adapthresh}} 
\end{figure}

For variant calling we now proceed as follows.  We assume that the majority
of the sites are non-variant, and only a smaller part is variant, with
$S_i>0$.  Thus, the observed distribution of $S_i$ will be a mixture
distribution, consisting of a null distribution corresponding to the
invariant sites and an alternative ``contamination'' component
corresponding to the variant sites.

In order to find an optimal adaptive cut-off separating the background from
potential variants we estimate the densities by fitting a natural spline
using Poisson-regression to the histogram, following the procedure
described by \citet{EfronTibshirani1996}.

Subsequently, we numerically
find the location with the minimum density and use it as threshold for
separating the two score populations. {Specifically, we fit
natural splines to the histogram for $S_i > 0$ and numerically determine the
local minimum.
If there are multiple minima the leftmost minimum is used.}
The corresponding threshold is
denoted by $S^*$.  

{In some cases there is no minimum in the histogram of the
  empirical scores.  In this case we use as fall-back solution the upper
  95\% quantile as threshold.  A missing minimum might indicate that the
  score model does not suffice to reliably call the variants.}

Once the threshold $S^*$ is
established, we declare all sites $S_i > S^*$ to be variant.
In \figcite{fig:adapthresh} this procedure is illustrated using data
from \textit{A.\ thaliana}.

{We note that by construction of the score $S_i$ we assume
  independence of the site characteristics.  However, in practice there
  will be correlation, and as alternative one may also consider a fully
  multivariate construction of the score $S_i$.  However, this is not
  without its own drawbacks, as the correlation among site characteristics
  may be hard to estimate reliably.  Moreover, as is well known from
  classification and ``naive Bayes'' analysis, independence models are
  typically rather robust and often even outperform more complex
  parameter-rich multivariate models.  }

\section*{Results and Discussion}

\subsection*{Simulation study}

{ To evaluate the reference implementation
  ``\texttt{haarz}'' of our adaptive model we compared it with the
    two frequently used SNV callers GATK \citep{MH+2010} and SAMtools
    \citep{LH+2009}}.  The precise command line settings are
  summarized in the Appendix.

We simulated next generation sequencing data for the human chromosome 21
using GemSIM (version 1.6)~\citep{MLT2012} with the default model and 
coverages ranging  from 10, 20, 30, 50, 100, to 200-fold. The simulated
content of the variant allele was either 0.2 or 0.5.  Simulated read length
was 100. For mapping we used the aligners that are recommended for each
method.  Specifically, we used BWA \citep{LD2009} to generate the
alignments for GATK and SAMtools. For the reference implementation of our
model we used \segemehl \citep{Hoffmann2009}.  After mapping and variant
calling we collected for each combination of coverage and variant allele
frequencies the number of false positives (FP), true positives (TP), false
negatives (FN), and true negatives (TN).  From this we computed the recall
(sensitivity) $SENS=TP/(TP+FN)$ and the positive predictive value
$PPV=TP/(FP+TP)$, { i.e. the true discovery rate.} For the
proposed data adaptive model we investigated the score distribution for all
12 experiments (\figcite{fig:distributions}).  Except for the combination
of low coverage (10x) and low variant allele content (20\%) we observe the
presence of two populations. The separability of these score populations
improves with increasing coverage and variant allele content.  In each
case, the minimum score for variant calls was automatically set to the
value where the density of scores $>0$ attains its first first local
minimum. Subsequently all positions with a score equal or greater were
called as SNV and compared to the other callers.

\begin{figure}
  \includegraphics[scale=0.9]{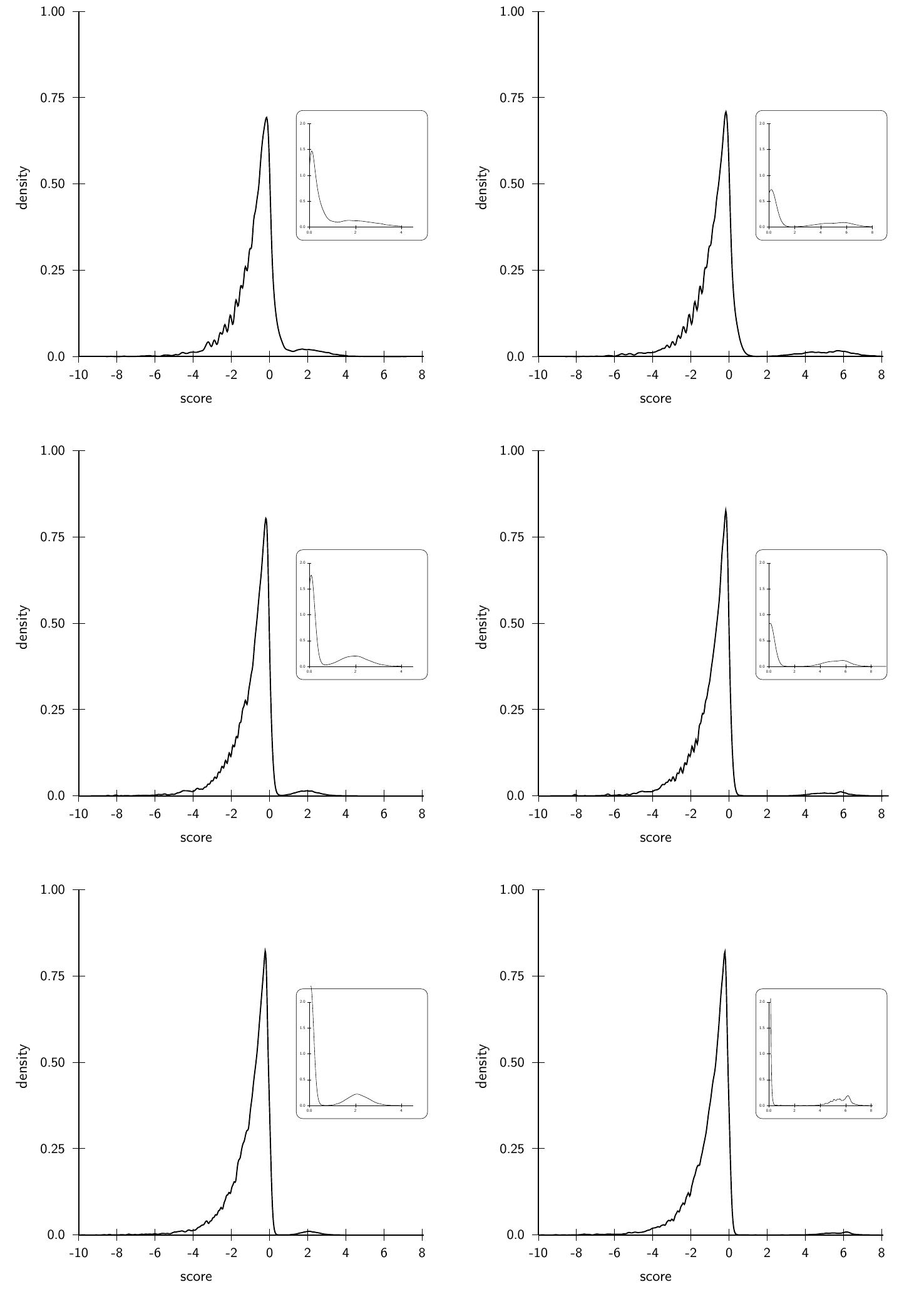}
  \caption{\textbf{Score distributions of simulated SNV data}. The left
    (right) column shows the score distributions of simulated SNVs with
    20\% (50\%) variant allele content for 20-fold (top), 50-fold (middle)
    and 100-fold(bottom) coverage. The insets show the density of scores
    $>0$. With the increase of coverage a population of scores $>0$ is
    clearly distinguishable from the background.\label{fig:distributions}} 
\end{figure}

All of the tested programs show a good recall and
positive predictive value in all 12 simulations. For low allele
contents in conjunction with low coverages, however, SAMtools attains
comparably low positive predictive values.  Surprisingly, after reaching a
maximum recall for the coverage of 100, the recall drops substantially for
coverage 200. For the simulations with 50\% allele content, all tools show
high recalls and good positive predictive values. Again, SAMtools achieves
only a comparably low positive predictive value for poorly covered SNVs
(\figcite{fig:roc2}). Except for the lowest coverage, all tools performed
well on these data sets.

\begin{figure} 
  \includegraphics[scale=0.9]{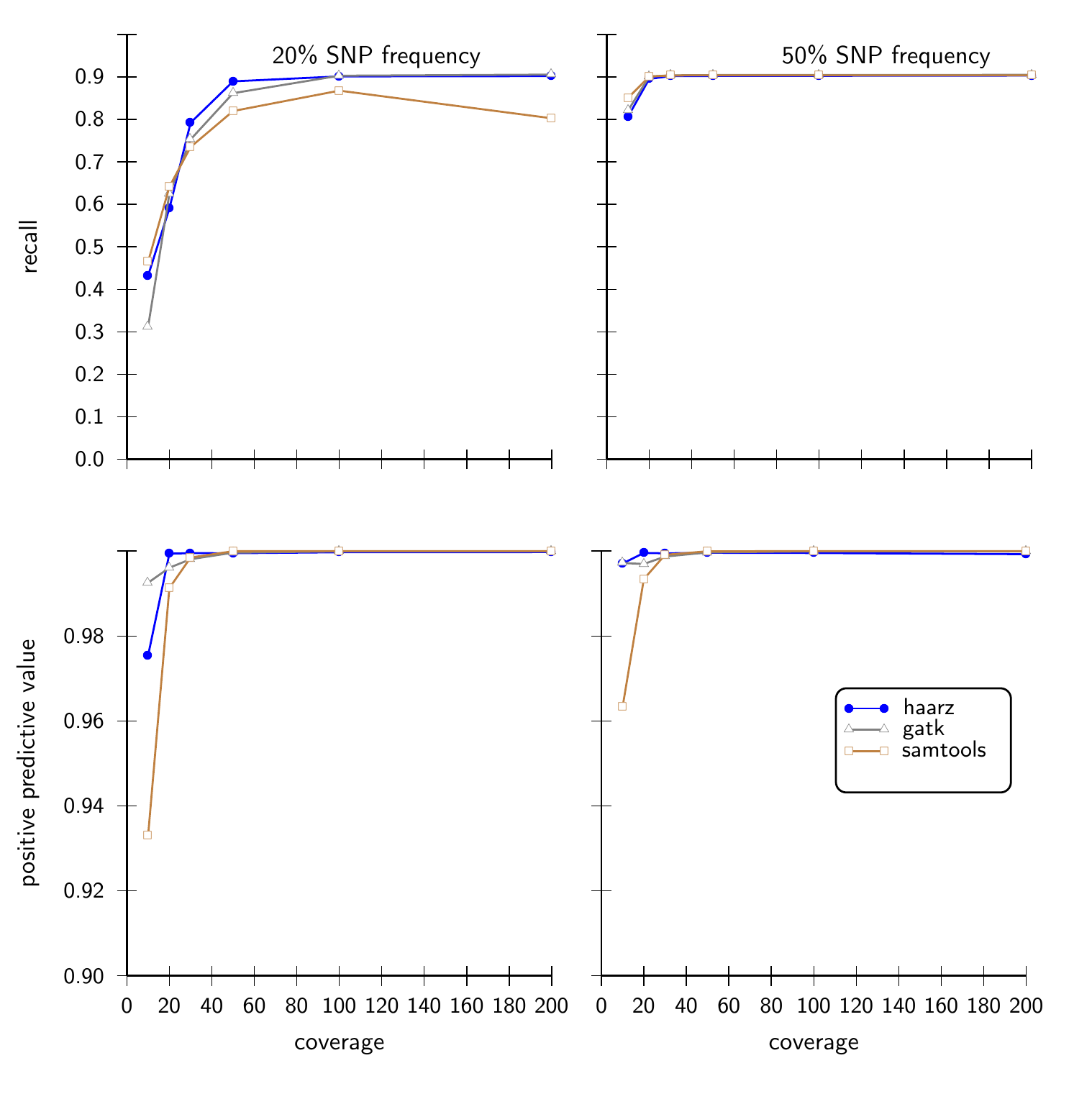} 
  \caption{\textbf{Statistical performance measures on simulated data
      sets.} The data adaptive model { implemented in
      \texttt{haarz}} was compared to SAMtools and GATK in terms of recall
    and positive predictive value. SNV calling was performed on twelve
    different data sets varying in the content of the variant allele (20\%
    and 50\%) as well as the simulated coverage (10-200). In all of these
    scenarios the data adaptive model is at par with both alternative
    callers.\label{fig:roc2}}
\end{figure}

{

In \figcite{fig:smallmaf} we show results for the challenging case of small minor
allele frequencies of 5\% and 10\%.  Our approach compares well in these rather
    difficult cases, in contrast to SAMtools and GATK. 
    For the low coverages, our algorithm does not find a clear
    cutoff and thus resorts to the 95\% criteria. Since there are very few
    sites with high scores, i.e. $S>0$, the recall is low and the positive
    predictive value is high. As soon as higher coverages are reached and a
    minimum is found, the recall is increases substantially. 
    We note that for low SNP frequencies
    in conjunction with low coverages the sample sizes for sampling site
    characteristics (default sample size: 100000; see Appendix) need to be increased
    to calculate the distribution of scores $S>0$.
}

\begin{figure} 
  \includegraphics[scale=0.9]{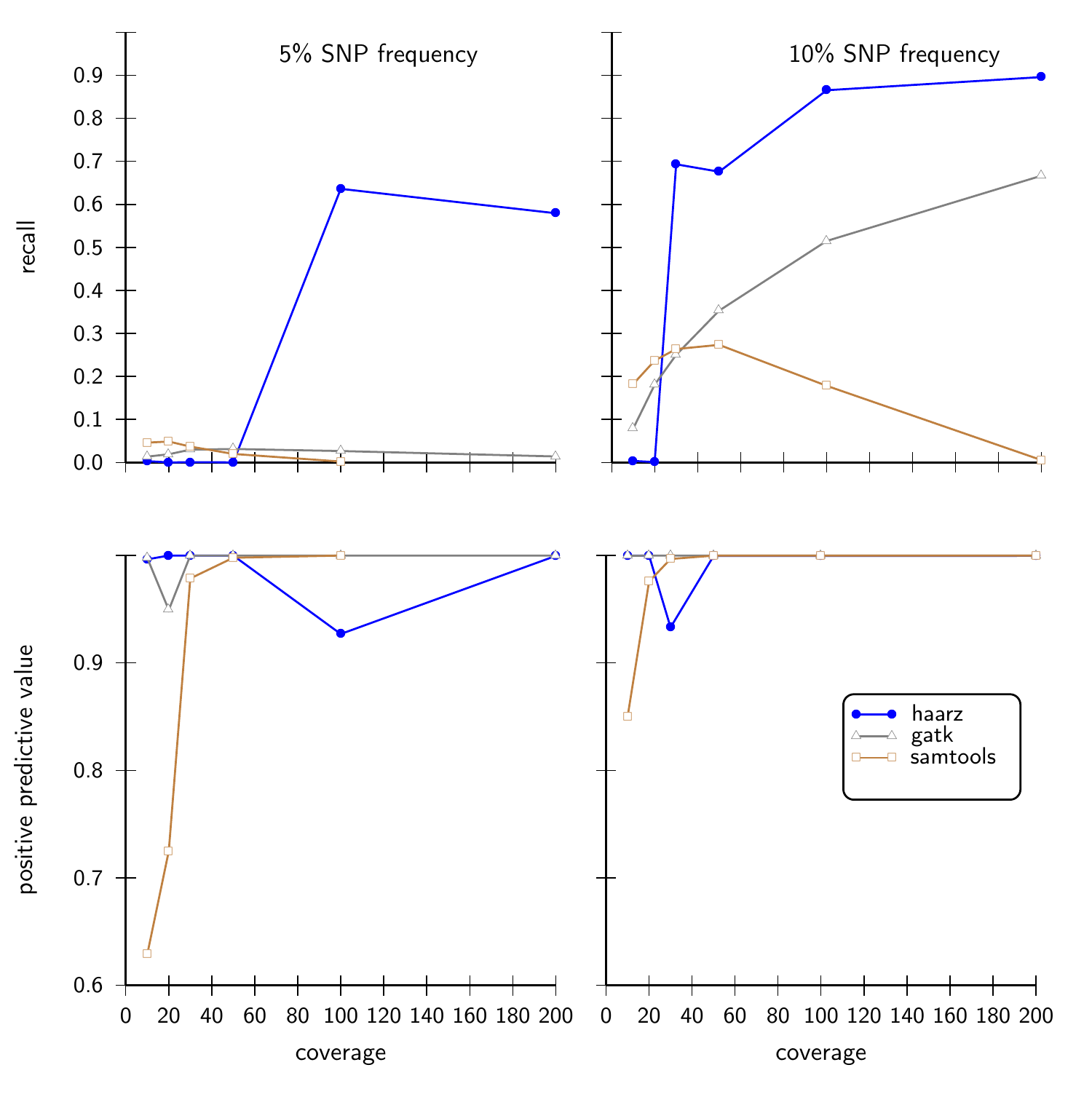} 
  \caption{\textbf{Statistical performance measures on simulated data
      sets with small variant allele frequency.} As in \figcite{fig:roc2}
      we compare our approach with SAMtools and GATK with 10\% and 5\% of
      minor allele frequencies. \label{fig:smallmaf}}
\end{figure}

\subsection*{Application to data sets}

The good overlap between the different methods in our simulation study as
well as the small number of false positives is in stark contrast to the
experience of greatly differing variant calls in real life data
\citep[e.g.][]{Rawe2013}. We therefore applied our model to diverse real
data from both diploid and haploid organisms. 

Paired end next generation sequencing data for \textit{Arabidopsis
  thaliana} (SRR519713), \textit{Escherichia coli} (ERR163894) and
\textit{Drosophila melanogaster} (SRR1177123) was downloaded under the
respective accession numbers from the Short Read Archive
(\url{www.ncbi.nlm.nih.gov/sra}). The \textit{Arabidopsis} data was aligned
to the reference genome version 10.5. With the data set SRR519713 we
obtained a coverage of $\sim$30-fold.  The \textit{E.\ coli} data set was
aligned to the reference genome \textit{E.\ coli k12} assembly v1.16.  With
ERR163894 we obtained a coverage of $\sim$60-fold. Finally, SRR1177123 was
aligned to the \textit{D.\ melanogaster} reference version dm3. The
coverage was $\sim$25-fold. For the alignments, calling and filtering we
used standard parameters. Precise settings are given in the Appendix.

The score distributions are shown in first line of
\figcite{fig:summary}. In the case of the plant \textit{A.\ thaliana} and
the procaryote \textit{E.\ coli}, a clear separation of two populations is
observable. On the other hand, the separation of the score populations in
\textit{D.\ melanogaster} data set is less pronounced.

In the lower part of \figcite{fig:summary} we show the concordance of
variant calls of three investigated methods for the three data sets.  For
the well separable cases, the number of calls made by the data adaptive
model is equal or higher as compared to the two competing callers.  For the
\textit{D.\ melanogaster} data set our approach is more conservative and
reports fewer variants. Most of these, however, are also found by SAMtools
and GATK. About 92\% percent of the calls from our model are also supported
by both of the other callers and only 4\% are not supported by any of the
two alternative approaches. From the score distributions for \textit{D.\
  melanogaster} it is clear that there is a large overlap of the two score
populations and hence the choice of $S^*$ necessarily depends on the
desired specificity and/or sensitivity.  
{ In the simulated data
(\figcite{fig:roc2} and \figcite{fig:smallmaf}) we see that
\texttt{haarz} generally achieve a high recall (sensitivity),
and at the same time offers a high positive predictive value, (PPV) i.e. low false discovery
rate. Thus, for the \textit{D.\ melanogaster} data many sites may be ambiguous
to call, and our tool will err on the conservative side to maintain a high PPV.
  }

\begin{figure}
  \begin{center}
    \includegraphics[width=\textwidth]{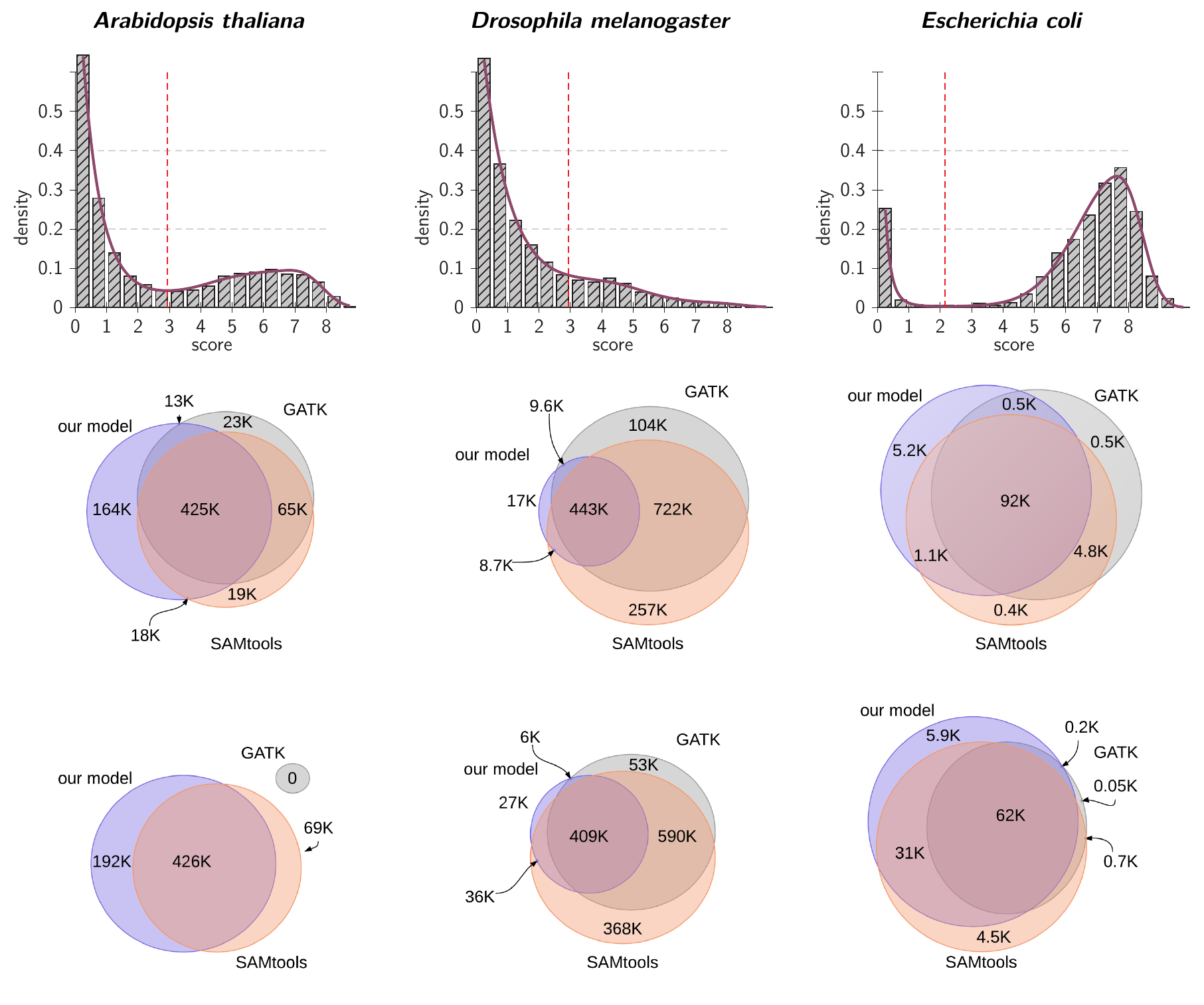}
  \end{center}
  \caption{\textbf{Score distributions and congruence of variant calls in
      next generation sequencing data}. While a clear separation of score
    populations is observable for the diploid \textit{A.\ thaliana} and the
    haploid \textit{E.\ coli}, only a shallow minimum can be observed in
    case of \textit{D.\ melanogaster}. Hence, in the latter case our
    model automatically adjusts to the data such that the calling appears
    to be more conservative: less than 4 percent of the calls are not
    supported by any caller and 92\% of its calls are supported both by
    GATK and SAMtools. On the other hand, our model calls more variants in
    the other two cases.\label{fig:summary} }
\end{figure}

\section*{Conclusions}

We have presented a data adaptive model for variant calling based on easily
accessible read characteristics, namely the log-likelihoods of nucleotide
qualities, relative read positions, alignment errors, multiple hits and the
mismatch rate at a position to obtain a score. With the exception of
nucleotide qualities, which are provided as input by the sequencing method,
all log-likelihoods are sampled from the data itself. We show that in
simulated as well as in real data sets this score gives rise to a mixture
distribution that distinguishes between true variants and noise.  A spline
fit to the overall marginal density allows us to determine a decision
threshold that optimally separates these score populations.  In the
simulated data we demonstrated that this simple model is at par with two of
the most commonly used probabilistic models for SNV calling methods in
terms of both sensitivity and specificity.

When applying our model to actual next-generation sequencing data, we
observe that the distributions of the scores vary significantly among the
different data sets. As expected, the clearest separation of the mixture
was obtained for the haploid \textit{E. coli} data set. In addition, the
small size of the genome and the absence of repetitive elements probably
improves the separability of the scores. The situation for the two diploid
genomes \textit{A.\ thaliana} and \textit{D.\ melanogaster} is
different. While both genomes have comparable sizes, the separability of
the score distributions varies strongly among these two data sets. While a
clear minimum can be found for the plant, the mixture in \textit{D.\
  melanogaster} appears to be more complicated.  In this case, by
construction our model selects a conservative decision threshold.  While
the number of calls is similar to the other probabilistic SNV callers in
the simulations as well as the next-generations data sets of the plant and
the bacteria, it is significantly reduced in the fruit fly data set.  These
differences indicate that the characteristics of next-generation data sets
have a strong impact on the success of variant calling. Furthermore, we
observe that, at least for the score proposed here, a significant
difference of the separability of the mixture distribution can be found
between simulated and real data. Thus, we argue that data adaptive
components could help to balance the trade-off between sensitivity and
specificity.

\section*{Acknowledgements}
This research was supported by LIFE (Leipzig Research Center for
Civilization Diseases), Leipzig University and the \emph{Deutsche
  Forschungs\-gemeinschaft} under the auspicies of SPP 1590 ``Probabilistic
Structures in Evolution'', proj.\ no.\ STA 850/14-1.  LIFE is funded by the
European Union, by the European Regional Development Fund (ERDF), the
European Social Fund (ESF) and by the Free State of Saxony within the
excellence initiative.
We thank the referees for their very helpful comments.

\newpage

\bibliographystyle{apalike}
\bibliography{snv}

\begin{thebibliography}{}

\bibitem[DePristo et~al., 2011]{DePristo2011}
DePristo, M.~A., Banks, E., Poplin, R., Garimella, K.~V., Maguire, J.~R.,
  Hartl, C., Philippakis, A.~A., del Angel, G., Rivas, M.~A., Hanna, M., et~al.
  (2011).
\newblock A framework for variation discovery and genotyping using
  next-generation dna sequencing data.
\newblock {\em Nature genetics}, 43(5):491--498.

\bibitem[Efron and Tibshirani, 1996]{EfronTibshirani1996}
Efron, B. and Tibshirani, R. (1996).
\newblock Using specially designed exponential families for density estimation.
\newblock {\em Annals Stat.}, 24:2431--2461.

\bibitem[Hoffmann et~al., 2009]{Hoffmann2009}
Hoffmann, S., Otto, C., Kurtz, S., Sharma, C.~M., Khaitovich, P., Vogel, J.,
  Stadler, P.~F., and {Hackerm\"uller}, J. (2009).
\newblock Fast mapping of short sequences with mismatches, insertions and
  deletions using index structures.
\newblock {\em PLOS Comp. Biol.}, 5:e1000502.

\bibitem[Li, 2011]{Li2011a}
Li, H. (2011).
\newblock A statistical framework for snp calling, mutation discovery,
  association mapping and population genetical parameter estimation from
  sequencing data.
\newblock {\em Bioinformatics}, 27(21):2987--2993.

\bibitem[Li and Durbin, 2009]{LD2009}
Li, H. and Durbin, R. (2009).
\newblock Fast and accurate short read alignment with {Burrows}-{Wheeler}
  transform.
\newblock {\em Bioinformatics}, 25:1754--1760.

\bibitem[Li et~al., 2009]{LH+2009}
Li, H., Handsaker, B., Wysoker, A., Fennell, T., Ruan, J., Horner, N., Martin,
  G., Abecasis, G., Durbin, R., and {1000 Genome Project Data Processing
  Subgroup} (2009).
\newblock The {Sequence} {Alignment/Map} format and {SAMtools}.
\newblock {\em Bioinformatics}, 25:2078--2079.

\bibitem[Liu et~al., 2013]{Liu2013}
Liu, X., Han, S., Wang, Z., Gelernter, J., and Yang, B.-Z. (2013).
\newblock {Variant Callers for Next-Generation Sequencing Data: A Comparison
  Study}.
\newblock {\em PLoS ONE}, 8(9):e75619+.

\bibitem[{McElroy} et~al., 2012]{MLT2012}
{McElroy}, K.~E., Luciani, F., and Thomas, T. (2012).
\newblock {GemSIM}: general, error-model based simulator of next-generation
  sequencing data.
\newblock {\em BMC Genomics}, 13:74.

\bibitem[{McKenna} et~al., 2010]{MH+2010}
{McKenna}, A., Hanna, M., Banks, E., Sivachenko, A., Cibulskis, K., Kernytsky,
  A., Garimella, K., Altshuler, D., Gabriel, S., Daly, M., and {DePristo1},
  M.~A. (2010).
\newblock The {Genome} {Analysis} {Toolkit}: A {MapReduce} framework for
  analyzing next-generation {DNA} sequencing data.
\newblock {\em Genome Res.}, 20:1297--1303.

\bibitem[O'Rawe et~al., 2013]{Rawe2013}
O'Rawe, J., Jiang, T., Sun, G., Wu, Y., Wang, W., Hu, J., Bodily, P., Tian, L.,
  Hakonarson, H., Johnson, W.~E., Wei, Z., Wang, K., and Lyon, G. (2013).
\newblock Low concordance of multiple variant-calling pipelines: practical
  implications for exome and genome sequencing.
\newblock {\em Genome Medicine}, 5(3):28.

\bibitem[Pabinger et~al., 2014]{Pabinger2013}
Pabinger, S., Dander, A., Fischer, M., Snajder, R., Sperk, M., Efremova, M.,
  Krabichler, B., Speicher, M.~R., Zschocke, J., and Trajanoski, Z. (2014).
\newblock A survey of tools for variant analysis of next-generation genome
  sequencing data.
\newblock {\em Briefings in Bioinformatics}, 15:256--278.

\bibitem[Xu et~al., 2014]{Xu2014}
Xu, H., DiCarlo, J., Satya, R., Peng, Q., and Wang, Y. (2014).
\newblock Comparison of somatic mutation calling methods in amplicon and whole
  exome sequence data.
\newblock {\em BMC Genomics}, 15(1):244.

\bibitem[Yu and Sun, 2013]{Yu2013}
Yu, X. and Sun, S. (2013).
\newblock Comparing a few snp calling algorithms using low-coverage sequencing
  data.
\newblock {\em BMC Bioinformatics}, 14(1):274.

\end{thebibliography}

\section*{Appendix}

\subsection*{Read simulation}
\noindent
For the simulation of reads and allele contents we used the program GemSIM (v. 1.6).
We simulated reads for the human chromosome 21 (hg19) with different coverages using
an Illumina specific error model (ill100v5\_s).

\footnotesize\begin{verbatim}
python GemReads.py -g simulatedsnps.txt -r chr21.fasta -m ill100v5_s.gzip \
-n <noofreads> -l 100 -q 64 -o <mysimulatedreads.fq>
\end{verbatim}
\normalsize

\subsection*{Benchmarks and command line parameters}
\noindent
For the benchmarks we have aligned the simulated as well as the real reads with
bwa and called the variants with SAMtools and GATK. For our own model
the reads were aligned using \segemehl. The command line parameters and
version numbers are given below.\\

\noindent
a) BWA v 0.6.2 
\footnotesize\begin{verbatim}
bwa aln <ref.fa>  <reads1.fq> > bwa_PE1.sai
bwa aln <ref.fa> <reads2.fq>  > bwa_PE2.sai
bwa sampe <ref.fa> bwa_PE1 bwa_PE2 <read1.fa> <read2.fa> > my.sam
\end{verbatim}
\normalsize

\noindent
b) GATK v 2.8.1 (GenomeAnalysisTK-2.8-1-g932cd3a) \\

\noindent
calling:
\footnotesize\begin{verbatim}
java -jar GenomeAnalysisTK.jar -T UnifiedGenotyper -R <ref.fa> \
     -I <sorted.bam> -o <calls.vcf>
\end{verbatim}
\normalsize
\noindent
filtering:
\footnotesize\begin{verbatim}
java -jar GenomeAnalysisTK.jar -T VariantFiltration -R <ref.fa> -V <calls.vcf> \ 
   --filterExpression "QD < 2.0 || FS > 60.0 || MQ < 40.0 || HaplotypeScore > 13.0 || \
   MQRankSum < -12.5 || ReadPosRankSum < -8.0" --filterName "filter" -o <calls.filtered.vcf>
\end{verbatim}
\normalsize

\noindent
c) SAMtools v 0.1.19\\

\noindent
calling:
\footnotesize\begin{verbatim}
samtools mpileup -uf <ref.fa> <sorted.bam> | bcftools view -bvcg - > <var.raw.bcf> 
bcftools view var.raw.bcf > <calls.vcf>
\end{verbatim}
\normalsize
\noindent
filtering:
\footnotesize\begin{verbatim}
varFilter -D100 > <calls.filter.vcf>
\end{verbatim}
\normalsize

\noindent
d) segemehl v 0.1.7\\

\noindent
\footnotesize\begin{verbatim}
segemehl.x -d <ref.fa> -i <ref.idx> -q SRR519713.fastq -D 0 > mysam.sam
\end{verbatim}
\normalsize
\noindent
obtaining site characteristics (written to sorted.haarz.idx):
\footnotesize\begin{verbatim}
haarz.x -d  <ref.fa> -q <sorted.sam.gz> -x <sorted.haarz.idx> -H -Q -M 2000
\end{verbatim}
\normalsize
\noindent
obtaining site characteristics for low variant allel frequencies:
\footnotesize\begin{verbatim}
haarz.x -d  <ref.fa> -q <sorted.sam.gz> -x <sorted.haarz.idx> -U 0 -X 0 -H -Q -M 2000
\end{verbatim}
\normalsize
\noindent
calling:
\footnotesize\begin{verbatim}
haarz.x -d <ref.fa> -q <sorted.sam.gz> -i <sorted.haarz.idx> -M 2000 -F
0.01 -Q -c > haarz.vcf
\end{verbatim}

\end{document}